\documentclass[fp,twocolumn]{jpsj3c}
\usepackage{txfonts}
\usepackage{color}

\title{Magnetic Structure and Dielectric State
in the Multiferroic Ca$_{2}$CoSi$_{2}$O$_{7}$}

\author{Minoru Soda$^{1}$, Shohei Hayashida$^{1}$, Toshiya Yoshida$^{1}$, 
Mitsuru Akaki$^{2}$, \\ Masayuki Hagiwara$^{2}$, 
Maxim Avdeev$^{3}$, Oksana Zaharko$^{4}$, and Takatsugu Masuda$^{1}$}
\inst{$^{1}$Neutron Science Laboratory, Institute for Solid State Physics, 
University of Tokyo, Tokai, Ibaraki 319-1106, Japan\\
$^{2}$Center for Advanced High Magnetic Field Science (AHMF), 
Graduate School of Science, Osaka University, Machikaneyama 1-1, 
Toyanaka, Osaka 560-0043, Japan\\
$^{3}$Bragg Institute, Australian Nuclear Science and Technology Organization, 
Lucas Heights, New South Wales 2234, Australia\\
$^{4}$Laboratory for Neutron Scattering and Imaging, 
Paul Scherrer Institut, CH-5232 Villigen PSI, Switzerland} 

\abst{Magnetic structure of the multiferroic Ca$_{2}$CoSi$_{2}$O$_{7}$ 
was determined by neutron diffraction techniques.
Combination of the polycrystalline and single-crystal samples experiments
revealed a collinear antiferromagnetic structure with 
the easy axis along $<$100$>$ directions.
The dielectric state was discussed in the  
framework of the spin-dependent $d$-$p$ hybridization mechanism, 
leading to the realization of the antiferroelectric structure. 
The origin of the magnetic anisotropy was discussed in comparison with the 
isostructural Ba$_{2}$CoGe$_{2}$O$_{7}$. 
}

\begin{document}
\maketitle

\section{Introduction}
The relationship between magnetism and dielectricity 
has attracted great interest in the field of condensed matter physics~ 
\cite{Fiebig,Eerenstein,Kimura,Cheong}. 
Among them multiferroics, 
in which the ferroelectricity is accompanied by the magnetic ordering, 
have been studied both experimentally \cite{Goto,Kimura2,Kitagawa} 
and theoretically \cite{Katsura,Mostovoy,Jia,Arima}. 
The microscopic mechanisms are explained  
by the spin-orbit coupling, which revealed the analytical relation between 
the spin structure and local electric polarization. 
For example, 
the multiferroic property in materials having cycloidal structures 
is explained by spin current mechanism~\cite{Katsura,Yamasaki,Cabrera,Soda-Z}, 
and that having a proper-screw structure is explained by 
the spin-dependent $d$-$p$ hybridization mechanism
\cite{Jia,Arima,K-Kimura,Soda-CCO}.
The change of the magnetic structure 
with temperature or magnetic field 
affects the ferroelectricity on the basis of the 
analytical relation\cite{Kimura-N,Sagayama,Soda-CCOH}. 

Recently, several multiferroic materials 
with a collinear structure 
have been studied~\cite{Kurumaji,Hayashida1,Hayashida2,Murakawa,Lorenz}. 
One example is a square-lattice antiferromagnet Ba$_{2}$CoGe$_{2}$O$_{7}$, 
the multiferroic properties of which have been explained by 
the $d$-$p$ hybridization mechanism~\cite{Murakawa}.
A large single-ion anisotropy $D$ of easy-plane type 
and a small spin-nematic interaction $J_p$, 
which is responsible for in-plane anisotropy, 
play an important role in the material~\cite{Soda,Soda2}. 
As for the spin dynamics,  
the large $D$ term induces an optical flat mode, 
which corresponds to an electric-field active mode confirmed by
electromagnetic wave absorption experiments~\cite{Kezsmarki}. 
The spin-nematic interaction is equivalent to the interaction between 
local electric polarizations, and the dielectric energy was probed 
in low energy range by an inelastic neutron scattering (INS)~\cite{Soda}. 
Furthermore, the direction of the magnetic moment in the $c$-plane 
was controlled 
by the electric field, 
which is quantitatively explained 
by the Hamiltonian including the dielectric energy 
estimated from the INS experiment.~\cite{Soda2}. 

Here, we focus on 
an isostructural material Ca$_{2}$CoSi$_{2}$O$_{7}$ 
in order to study physical properties of the material 
having a different set of $D$, $J_p$, 
and additional orthorhombic anisotropy $E$. 
The Ca$_{2}$CoSi$_{2}$O$_{7}$ 
has a two-dimensional network of 
CoO$_{4}$ and SiO$_{4}$ tetrahedra as shown in Fig. 1(a). 
Co$^{2+}$ ions have spin $S$ = 3/2 
and form a square lattice. 
Below $T_{\mathrm{N}}$ = 6 K, 
a magnetic order appears, 
and the electric polarization is induced 
by the magnetic field $H$ \cite{Akaki1,Akaki2}. 

\begin{figure}[b]
\begin{center}\leavevmode
\includegraphics[width=7 cm]{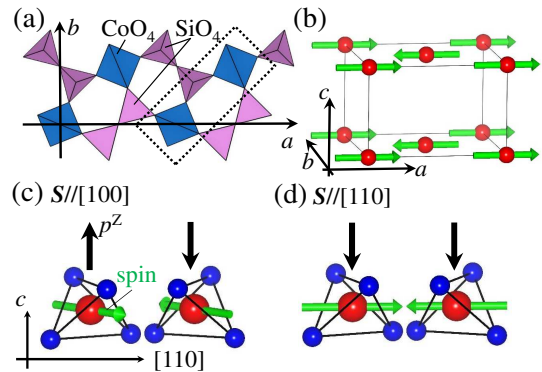}
\caption{
(color online) 
(a) Two-dimensional network of CoO$_{4}$ and SiO$_{4}$ 
tetrahedra in Ca$_{2}$CoSi$_{2}$O$_{7}$. 
(b) Schematic magnetic structure of Ca$_{2}$CoSi$_{2}$O$_{7}$. 
(c,d) Dielectric state 
of CoO$_{4}$ tetrahedron 
for the magnetic structures 
with the easy axes along (c) $<$100$>$ and (d) $<$110$>$. 
Green arrows indicate the Co$^{2+}$ spin ($S$), 
and black arrows indicate 
the local electric polarization along the $c$-axis ($p^Z$). 
}
\label{Fig. 1}
\end{center}
\end{figure}

In the present study, neutron diffraction measurements have been performed 
to clarify the magnetic structure of Ca$_{2}$CoSi$_{2}$O$_{7}$.
The determined structure was a collinear antiferromagnetic one 
with the easy axis along $\verb|<|100\verb|>|$ directions.  
Consideration on the multiferroic property in the framework of 
the spin-dependent $d$-$p$ hybridization mechanism led to the 
same dielectric state as that of the isostructural Ba$_{2}$CoGe$_{2}$O$_{7}$. 
The discussion on the magnetization data on the basis 
of the determined magnetic structure 
revealed that the origin of the in-plane anisotropy 
that dominates the multiferroic property is 
orthorhombic anisotropy $E$. 

\section{Experimental Details}
The single crystals of  Ca$_{2}$CoSi$_{2}$O$_{7}$ were grown 
by the floating-zone method~\cite{Akaki1}. 
The polycrystalline sample was obtained 
by pulverizing the single crystal. 
The magnetization was measured by a commercial SQUID magnetometer. 
Powder neutron diffraction experiments were performed 
at ECHIDNA spectrometer installed at OPAL in ANSTO, Australia. 
The used neutron wavelength with Ge (331) monochromator was 2.4395 \AA. 
The single-crystal neutron diffraction experiment 
was carried out using 
TriCS spectrometer installed at SINQ in PSI, Switzerland. 
The used neutron wavelength with Ge (311) monochromator was 1.18 \AA. 
The scattering plane was the $a$-$b$ plane, and
throughout this paper, we use the tetragonal unit cell with the space group 
$P\overline{4}2_1m$
where the lattice constants are $a$ = $b$ = 7.844 \AA\ and $c$ = 5.027 \AA. 
The sample was cooled by using a liquid-helium cryostat. 

\section{Results and Analysis}
\subsection{Magnetization}
Figure 2 shows the magnetic field ($H$) dependence 
of the magnetization ($M$) on single-crystal measured at 1.8 K 
for {\itshape\bfseries{H}}$\parallel$[100] 
and {\itshape\bfseries{H}}$\parallel$[110]. 
Hysteresis induced by the weak ferromagnetic component 
is observed below $H$ = 1 T. 
The magnetization for {\itshape\bfseries{H}}$\parallel$[100] 
shows nonlinear behaviour at about 3.5 T, 
while the magnetization for {\itshape\bfseries{H}}$\parallel$[110] 
indicates almost linear behaviour at 1 T $<$ $H$ $<$ 5 T. 
These are in contrast 
with the isostructural material Ba$_{2}$CoGe$_{2}$O$_{7}$; 
the $M$ in the field {\itshape\bfseries{H}}$\parallel$[110] is nonlinear and 
the $M$ in the field {\itshape\bfseries{H}}$\parallel$[100] is linear
~\cite{Soda}.  
The inset of Fig.~2 depicts the low field region of magnetization.  
This is consistent with the previous study of the magnetization~\cite{Akaki2}. 

\begin{figure}[t]
\begin{center}\leavevmode
\includegraphics[bb=10 430 400 800, width=7.3 cm, clip]{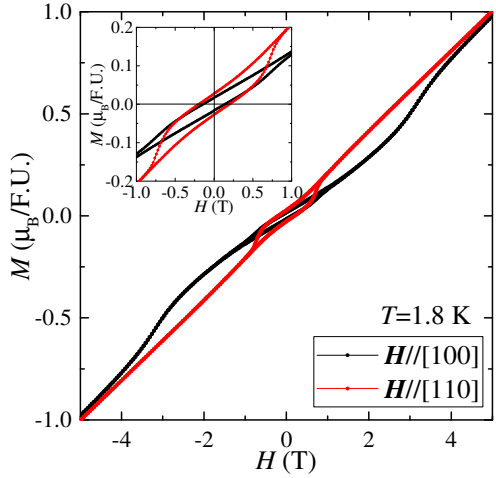}
\caption{
(color online) 
Magnetic-field dependence of magnetization on single crystal 
measured at $T$=1.8 K 
for {\itshape\bfseries{H}}$\parallel$[100] and 
{\itshape\bfseries{H}}$\parallel$[110].
Inset shows 
the low-field region of the magnetization.
}
\label{Fig. 2}
\end{center}
\end{figure}

\subsection{Powder Neutron Diffraction}
\begin{figure}[t]
\begin{center}\leavevmode
\includegraphics[width=7.5 cm]{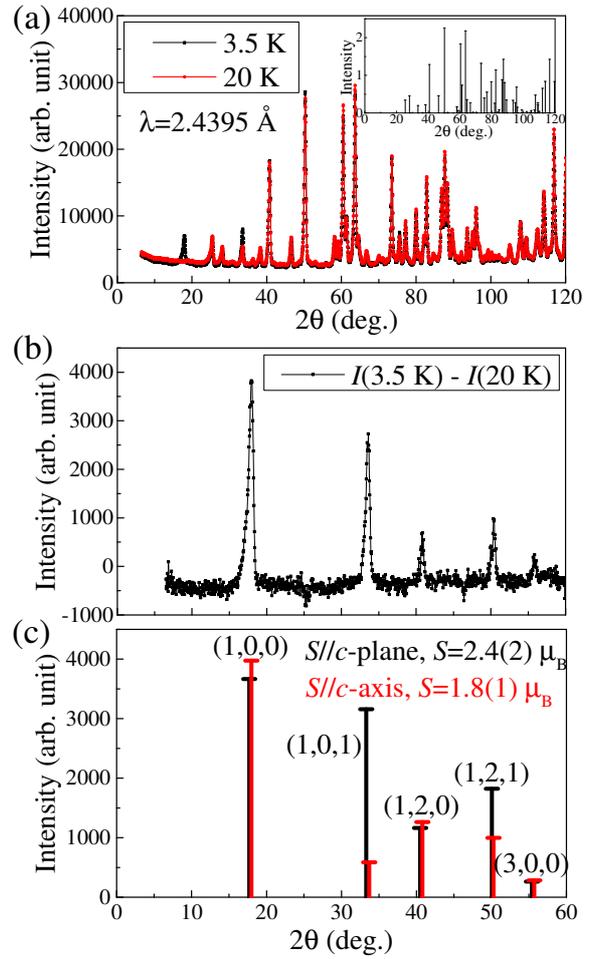}
\caption{
(color online) 
(a) Neutron-powder-diffraction patterns measured 
at 3.5 K and 20 K. 
Inset shows the calculated intensities of nuclear reflections 
with the space group $P\overline{4}2_1m$. 
(b) Magnetic intensities obtained by subtracting the data at 20 K 
from that at 3.5 K. 
(c) Magnetic intensities calculated for the collinear-magnetic structures 
with the easy axis in the $c$-plane and along the $c$-axis. 
}
\label{Fig. 3}
\end{center}
\end{figure}

The main panel in Fig. 3(a) shows the neutron powder-diffraction 
patterns measured at 3.5 K and 20 K. 
The inset shows the calculation of nuclear reflections on the basis of 
the averaged crystal structure 
with the space group $P\overline{4}2_1m$~\cite{Kusaka}. 
Figure 3(b) shows the magnetic intensities 
obtained by subtracting the data at 20 K from that at 3.5 K. 
Several magnetic reflections are observed in the low angle region, 
and all the magnetic peaks are indexed by a propagation vector (0,0,0). 
The obtained vector 
coincides with that in Ba$_{2}$CoGe$_{2}$O$_{7}$ \cite{Zheludev}. 
In Fig. 3(c), the magnetic intensities are calculated for two models; 
the collinear antiferromagnetic structure with the spins 
confined in the $c$-plane 
and that with the spins directed along the $c$-axis. 
It is noted that the direction of the magnetic moment 
in the $c$-plane can not be determined 
by a powder neutron-diffraction technique 
because the technique is not sensitive to magnetic domains. 
In the calculation, the isotropic magnetic form factor 
of Co$^{3+}$ was used~\cite{formfactor}.  
The estimated magnitude of the Co-magnetic moment in the former 
model is 2.4(2) $\mu_{\rm B}$ and that in the latter is 1.8(1) $\mu_{\rm B}$. 
The former model reproduces the experiment better, and we found that 
Ca$_{2}$CoSi$_{2}$O$_{7}$ has a collinear antiferromagnetic structure 
with the moments confined in the $c$-plane.  
 
\subsection{Single Crystal Experiment}
The main panel and the inset of Fig.~4(a) 
show the temperature ($T$) dependence of the integrated intensity 
and the typical $\omega$-scan profiles
at {\itshape\bfseries{Q}} = (0,1,0), respectively. 
The small intensity at the {\itshape\bfseries{Q}}-point 
exists above $T_N$, which is originated from the nuclear reflection. 
With decreasing $T$, a large intensity component due to 
magnetic ordering appears below $T_N$. 

\begin{figure}[t]
\begin{center}\leavevmode
\includegraphics[width=7.5 cm]{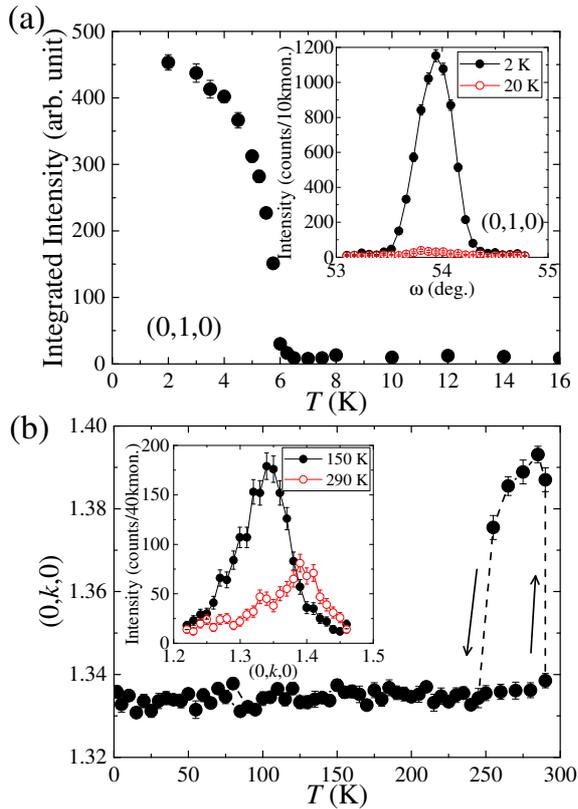}
\caption{
(color online) 
(a) $T$-dependence of the integrated intensity 
at the {\itshape\bfseries{Q}}-point (0,1,0).
Inset shows the typical $\omega$-scan profiles. 
(b) $T$-dependence of the $k$-value of the superlattice reflection 
at (0,$k$,0). 
Inset shows the typical profiles along $k$-direction. 
}
\label{Fig. 4}
\end{center}
\end{figure}

Nuclear superlattice reflections are observed 
at ($h_0$$\pm$$\delta$,$k_0$,$l_0$), ($h_0$,$k_0$$\pm$$\delta$,$l_0$), 
and ($h_0$$\pm$$\delta$,$k_0$$\pm$$\delta$,$l_0$) 
($h_0$, $k_0$, and $l_0$ = integer number, and $\delta$ = noninteger number).
The inset of Fig. 4(b) shows 
the typical profiles of the superlattice reflection 
along $k$-direction at (0,$k$,0), where $k=1-\delta$. 
It is clear that the superlattice reflection changes with the temperature. 
The $T$-dependence of the $k$-value 
of the superlattice reflection 
at (0,$k$,0) is shown in Fig. 4(b). 
In the wide $T$ region, 
the nuclear superlattice reflections with $\delta$=1/3 are observed. 
At around 250 K, 
the $\delta$-value of the superlattice reflection changes 
with hysteresis, 
suggesting that the structural transition takes place 
at this temperature~\cite{Kusaka2}. 

The observed superlattice peaks are explained by 
a 3$\times$3$\times$1 supercell 
with the space groups $P$$2_1$$2_1$2~\cite{Sazonov}. 
The main purpose of our study is, however, the magnetic structure of 
Ca$_{2}$CoSi$_{2}$O$_{7}$ instead of detailed crystal structure analysis. 
Since the nuclear intensity is used only for the scaling factor 
to estimate the 
magnitude of the magnetic moment in the magnetic structure analysis, 
the use of the averaged crystal structure is reasonable. 
Also, the nuclear reflection observed at (0,1,0) is 
prohibited in $P$$2_1$$2_1$2 and the symmetry of the crystal 
is further reduced. 
In the present study, we analyzed the magnetic structure 
using the averaged space group $P\overline{4}2_1m$~\cite{Kusaka}. 

The magnetic diffraction measurements 
on the Ca$_{2}$CoSi$_{2}$O$_{7}$ single crystal were carried out 
in order to clarify the direction of the magnetic moment in the $c$-plane. 
At 20 K and 2 K, the neutron intensities were measured 
at the {\itshape\bfseries{Q}}-points equivalent to (1,2,0). 
The magnetic intensities 
were obtained by subtracting the integrated intensity at 20 K 
from that at 2 K. 
In order to avoid the neutron absorption effect, 
the magnetic intensities were divided by the nuclear intensities 
at each {\itshape\bfseries{Q}}-point. 
Then, the magnetic intensities at the equivalent {\itshape\bfseries{Q}}-points 
can be compared relatively and accurately 
on the assumption 
that the nuclear intensities at the equivalent {\itshape\bfseries{Q}}-points 
are equal. 
The average of the obtained magnetic intensities 
is normalized to 1.0,  
and the magnetic intensity at each {\itshape\bfseries{Q}}-point 
is plotted as a function of the angle 
between the scattering vector {\itshape\bfseries{Q}} and the $a$-axis 
in Fig.~5(a). 
This result shows that the magnetic intensity has the angle dependence 
in the $c$-plane. 

The magnetic neutron scattering is caused 
by the component of the moment perpendicular 
to the scattering vector {\itshape\bfseries{Q}}, and thus 
the angle dependence of the intensity 
at equivalent {\itshape\bfseries{Q}}-points 
is expressed by the polarization factor 2(1$-$$S_{Q}^2$/$S^2$). 
Here the $S_{Q}$ is the component of the magnetic moment parallel to 
the scattering vector {\itshape\bfseries{Q}}. 
Using the formula, the magnetic intensities of the collinear structure 
with the spins aligned along [100] and [010] directions 
were calculated; the former is indicated by the blue curve 
and the latter is by the red curve in Fig. 5(a). 
In order to explain the angle dependence, 
we need to consider the imbalance of different magnetic domains. 
The volume percentages of the domains 
with the easy axis along [100] and [010] directions 
are estimated to be 60 and 40\%, respectively, 
as indicated by the black curve in Fig. 5(a). 
In contrast, the assumed domains of the magnetic structure 
with the easy axis along $\verb|<|110\verb|>|$ directions, 
in which the intensity curve shifts 
by 45 degree from that for [100] and [010] directions, 
do not reproduce the data. 
Accordingly, the magnetic structure 
with the easy axis along $\verb|<|100\verb|>|$ directions (Fig.~1(b)) 
is realized. 

The collected 69 magnetic reflections 
were analyzed by using the above magnetic structure. 
The intensities were obtained 
by subtracting the integrated intensity at 20 K 
from that at 2 K. 
We fixed the volume ratio between the domains 
with the easy axis along [100] and [010] directions as $60 : 40$, 
and we calculated the magnetic cross section $I$. 
Figure 5(b) shows the observed $I_{\rm obs.}$ 
against the model calculation $I_{\rm cal.}$. 
The magnitude of the aligned magnetic moment 
is 3.1 ($\pm$0.2) $\mu_B$ which is close to the full moment of Co$^{3+}$ ion. 
The observed magnetic intensities are reasonably reproduced 
by the magnetic structure shown in Fig. 1(b). 

Although the canted antiferromagnetic structure is expected from 
the magnetization curve in Fig.~2, 
the small ferromagnetic component could not be estimated 
in our experimental resolution. 
The residual magnetization for
{\itshape\bfseries{H}}$\parallel$[100] 
is about 0.02 $\mu_B$/Co$^{3+}$ as shown in the inset of Fig. 2. 
One of the magnetic domains has the easy axis along [100] 
and the ferromagnetic component along [010], 
and another one of the magnetic domains has the easy axis along [010] 
and the ferromagnetic component along [100]. 
The actual magnitude of the ferromagnetic component per one Co ion 
is evaluated to be about 0.04 $\mu_B$ 
by considering the magnetic domains. 
Then, the canted angle from $a$- or $b$-axis is about 1 degree. 
As in Ba$_{2}$CoGe$_{2}$O$_{7}$ \cite{Soda}, 
this small canting of the neighboring magnetic moments is attributed to  
the Dzyaloshinskii-Moriya interaction.
\begin{figure}[t]
\begin{center}\leavevmode
\includegraphics[width=7.5 cm]{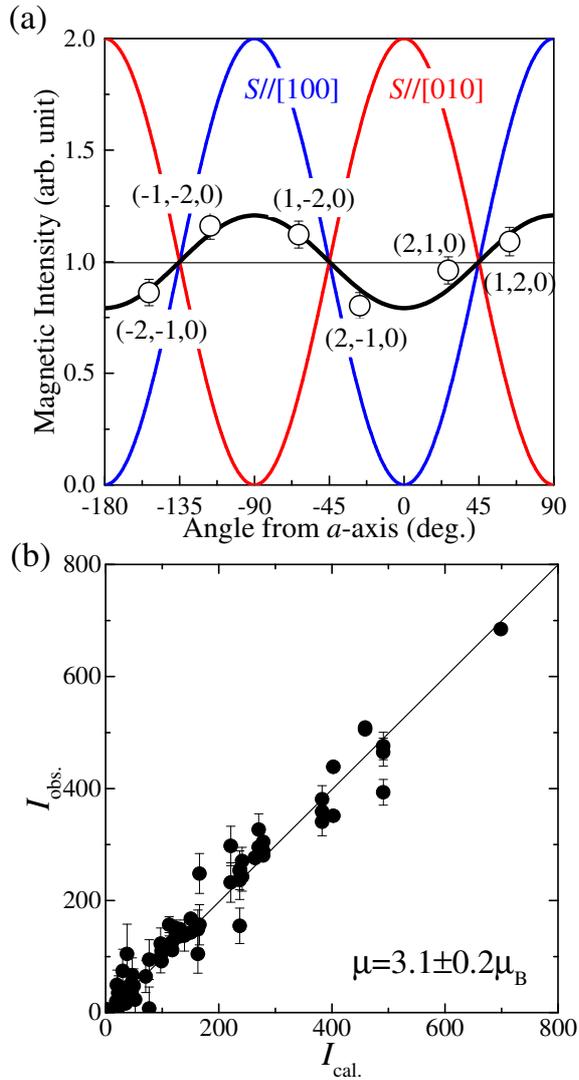}
\caption{
(color online) 
(a) Magnetic intensities 
at the {\itshape\bfseries{Q}} points equivalent
to (1,2,0) against the angle from the $a$-axis. 
Calculated intensities for the magnetic structures 
with the magnetic moments along [100] (red line) and [010] (blue line). 
Black line shows the angle dependence of the magnetic intensity 
calculated with the imbalance of the magnetic domains 
with the easy axis along [100] and [010] directions. 
(b) Magnetic cross sections obtained by subtracting the integrated intensity 
at 20 K from that at 2 K 
are shown against those of the model calculation.
}
\label{Fig. 5}
\end{center}
\end{figure}

\section{Discussion}
We discuss the relationship 
between the magnetic structure and 
the multiferroic properties of Ca$_{2}$CoSi$_{2}$O$_{7}$. 
Since the collinear antiferromagnetic structure was identified, 
the ferroelectric properties \cite{Akaki1,Akaki2} 
can be explained neither by 
the spin-current model \cite{Katsura,Yamasaki} nor 
the double exchange mechanism \cite{Sergienko,Lorenz}. 
Accordingly, 
we consider the spin-dependent $d$-$p$ hybridization mechanism \cite{Arima}
as in Ba$_{2}$CoGe$_{2}$O$_{7}$ \cite{Murakawa}. 
In this mechanism, 
the magnitude and direction of the local electric polarization depend on  
the direction of the local Co$^{3+}$ spin 
and the directions of Co-O bondings in CoO$_{4}$ tetrahedra. 
This relation is expressed as 
\begin{eqnarray}
\mbox{\boldmath $p$}\propto\sum_{i}(\mbox{\boldmath $S$}\cdot
\mbox{\boldmath $e$}_{i})^{2}\mbox{\boldmath $e$}_{i},
\end{eqnarray}
where {\itshape\bfseries{e}}$_{i}$ is the vector 
connecting the Co and the $i$-th O ions~\cite{Jia,Arima,Murakawa}. 
The electric polarization $\mbox{\boldmath$p$}$ 
points to the $c$-axis from the relationship of Eq. (1) 
because the Co magnetic moments are confined in the $c$-plane.

In considering Eq. (1) and the crystal structure shown in Fig. 1(a), 
the dielectric state depends on 
the direction of the Co magnetic moment~\cite{Soda,Soda2}. 
The magnetic structure in Fig. 1(b) 
induces the antiferroelectric structure as shown in Fig. 1(c), 
while the collinear structure 
with the easy axis along $\verb|<|110\verb|>|$ directions 
induces the ferroelectric state as shown in Fig. 1(d). 
In Ca$_{2}$CoSi$_{2}$O$_{7}$, thus, 
the antiferroelectric structure must be realized. 

Now let us discuss the magnetization data in Fig. 2 based on the collinear 
structure identified by the present neutron diffraction. 
Assuming that the dominant in-plane anisotropy in Ca$_{2}$CoSi$_{2}$O$_{7}$ is 
orthorhombic anisotropy and the origin is 
single-ion type $E((S^{[100]})^2-(S^{[010]})^2)$ with negative $E$, 
the spin flop transition is present 
in the field {\itshape\bfseries{H}}$\parallel$[100] and 
it is absent in {\itshape\bfseries{H}}$\perp$[100]. 
In the field {\itshape\bfseries{H}}$\parallel$[110] 
which is tilted by 45$^{\circ}$ from the easy axis, 
the spin flop anomaly is strongly suppressed and smeared. 
This behavior is qualitatively consistent 
with the the magnetization data exhibiting 
the anomaly at $H \sim 3.5$ T in {\itshape\bfseries{H}}$\parallel$[100] 
and the absence of anomaly in {\itshape\bfseries{H}}$\parallel$[110], 
except hysteretic behavior in $|H| \lesssim 1$ T. 
The assumption is, thus, reasonable and the single-ion anisotropy $E$ is the 
dominant origin for the in-plane anisotropy instead of  
spin-nematic interaction $J_p$ originated  
from the interaction between local electric polarization.  

This contrasts with the isostructural Ba$_{2}$CoGe$_{2}$O$_{7}$ which 
has the same magnetic structure 
and electric polarization structure~\cite{Soda,Soda2}. 
The origin of the anisotropy in Ba$_{2}$CoGe$_{2}$O$_{7}$ 
is the spin-nematic interaction 
and the anisotropy has four-fold rotational symmetry. 
This leads to different type of spin flop; 
the spins flop from [100] or [010] to [1$\bar{1}$0] at $H$=0.35 T 
in {\itshape\bfseries{H}}$\parallel$[110]~\cite{Soda}. 
In fact the symmetry of the Ba$_{2}$CoGe$_{2}$O$_{7}$ is also lowered
by slight lattice distortion~\cite{Hutanu} 
and orthorhombic anisotropy in the $c$-plane is allowed. 
The spin nematic interaction is, however, dominant and exotic type of 
spin flop transition was exhibited. 

While the magnetization in the low-$H$ region was explained by 
the in-plane anisotropy, distinct anomaly 
for {\itshape\bfseries{H}}$\parallel$[110] at 11 T 
and non-trivial magnetization plateau 
for {\itshape\bfseries{H}}$\parallel$$c$-axis above 
18 T~\cite{Akaki2} cannot be explained. 
Small energy scale of the single-ion anisotropy $D$ is suggested by 
the magnetization measurements~\cite{Akaki2,Yi,Penc}, and 
proximity of the $D$ and the exchange interaction would be the origin for 
the anomalous behavior in a high field. 
Detailed study on spin dynamics will be the key to the understanding 
of the relationship 
among the magnetization, the single-ion anisotropy, 
and the exchange interaction. 

\section{Conclusions}
Combination of neutron diffraction experiments on polycrystalline and 
single-crystal samples has revealed magnetic structure of the  
multiferroic Ca$_{2}$CoSi$_{2}$O$_{7}$. 
The former experiment has shown the antiferromagnetic collinear structure 
in the $c$-plane. 
The latter experiment including precise measurements on several equivalent 
Bragg peaks has shown that the spin directions are along $\verb|<|100\verb|>|$. 
Antiferroelectric structure has been derived from the discussion on the 
multiferroic property on the basis of  the spin-dependent $d$-$p$ hybridization mechanism.
The origin of the magnetic anisotropy that determines the multiferroic 
property has been discussed, 
and the orthorhombic anisotropy $E$ is dominant in 
the in-plane anisotropy. 

\begin{acknowledgments}
Travel expenses to SINQ for the neutron experiments
were supported by General User Program 
for Neutron Scattering Experiments, 
Institute for Solid State Physics, 
The University of Tokyo (proposal no. 15519). 
This work was supported in part by KAKENHI 
(15K05123 and 15K05145) from MEXT, Japan.
\end{acknowledgments}

\end{document}